\documentclass[aps,prx,amsmath,superscriptaddress,twocolumn,bibnotes,longbibliography,floatfix]{revtex4-2}
\usepackage{amsmath}
\usepackage{amsfonts}
\usepackage{graphicx}
\usepackage{color}
\usepackage{dsfont}
\usepackage{verbatim}
\usepackage{mathtools}
\usepackage[normalem]{ulem}
\usepackage{comment}
\usepackage{stackrel}
\usepackage{bm}
\usepackage{booktabs}
\usepackage{comment}

\usepackage[colorlinks=true, 
linkcolor=blue, 
urlcolor=blue,
citecolor=blue]{hyperref}
\usepackage{orcidlink}
\usepackage[linesnumbered]{algorithm2e}

\RestyleAlgo{ruled}
\SetKwComment{Comment}{/* }{ */}

\definecolor{myblue}{rgb}{.93, .93, 1}

\newcommand{\bsub}{\begin{subequations}}
	\newcommand{\esub}{\end{subequations}}

\newcommand{\vex}[1]{\bm{\mathrm{#1}}}

\usepackage{xcolor}
\usepackage{soul}

\begin{document}
	
	\title{Intravalley spin-polarized superconductivity in rhombohedral tetralayer graphene}
	
	\author{Yang-Zhi~Chou~\orcidlink{0000-0001-7955-0918}}\email{yzchou@umd.edu}
	\affiliation{Condensed Matter Theory Center and Joint Quantum Institute, Department of Physics, University of Maryland, College Park, Maryland 20742, USA}
	
	\author{Jihang Zhu}
	\email{jizhu223@gmail.com}
	\affiliation{Condensed Matter Theory Center and Joint Quantum Institute, Department of Physics, University of Maryland,
		College Park, Maryland 20742, USA}

	\author{Sankar Das~Sarma}
	\affiliation{Condensed Matter Theory Center and Joint Quantum Institute, Department of Physics, University of Maryland, College Park, Maryland 20742, USA}
	\affiliation{Kavli Institute for Theoretical Physics, University of California, Santa Barbara, California 93106, USA}
	
	\date{\today}
	
	\begin{abstract}
		We study the intravalley spin-polarized superconductivity in rhombohedral tetralayer graphene, which has been discovered experimentally in Han $et$ $al$ arXiv:2408.15233. We construct a minimal model for the intravalley spin-polarized superconductivity, assuming a simplified anisotropic interaction that depends only on the angle between the incoming and outgoing momenta. Despite the absence of \textit{Fermi surface nesting}, we show that superconductivity can emerge near the Van Hove singularity with the maximal $T_c$ near a bifurcation point of the peaks in the density of states. We identify the $p+ip$, $h+ih$, and the nodal $f$-wave pairings as the possible states, which are all pair density wave orders due to the intravalley nature. Furthermore, these pair density wave orders require a finite attractive threshold for superconductivity, resulting in {a narrow stripe shape of superconducting region}, consistent with experimental findings. We point out that the Kohn-Luttinger mechanism is a plausible explanation with a dominant $p+ip$ pairing. The possibility of realizing intravalley spin-polarized superconductivity in other rhombohedral graphene systems is also discussed.
	\end{abstract}
	
	\maketitle
	
\section{Introduction}
    
	Superconductivity (SC) is one of the most important and intriguing quantum phenomena in condensed matter and material physics. Since the initial discovery of SC in the magic-angle twisted bilayer graphene \cite{CaoY2018a}, observable two-dimensional (2D) SC ($T_c>20$ mK) has been reported in various twisted and untwisted van der Waals multilayer systems \cite{BalentsL2020,AndreiEY2021} (e.g., twisted multilayer graphene \cite{YankowitzM2019,LuX2019,ChoiY2019,AroraHS2020,ParkJM2021,HaoZ2021,CaoY2021,OhM2021,ParkJM2022,ZhangY2022,SuR2023}, twisted bilayer WSe$_2$ \cite{XiaY2024,GuoY2024}, Bernal bilayer graphene \cite{ZhouH2022,ZhangY2023,HolleisL2024,LiC2024a,ZhangY2024}, and rhombohedral trilayer \cite{ZhouH2021a,YangJ2024,PattersonCL2024} and tetralayer graphene \cite{ChoiY2024}).  
	Notably, these untwisted graphene multilayers are promising systems for studying unconventional SC because of the ability to control the electronic band structures through the displacement field and the lower disorder nature.

	\begin{figure}[t!]
		\includegraphics[width=0.41\textwidth]{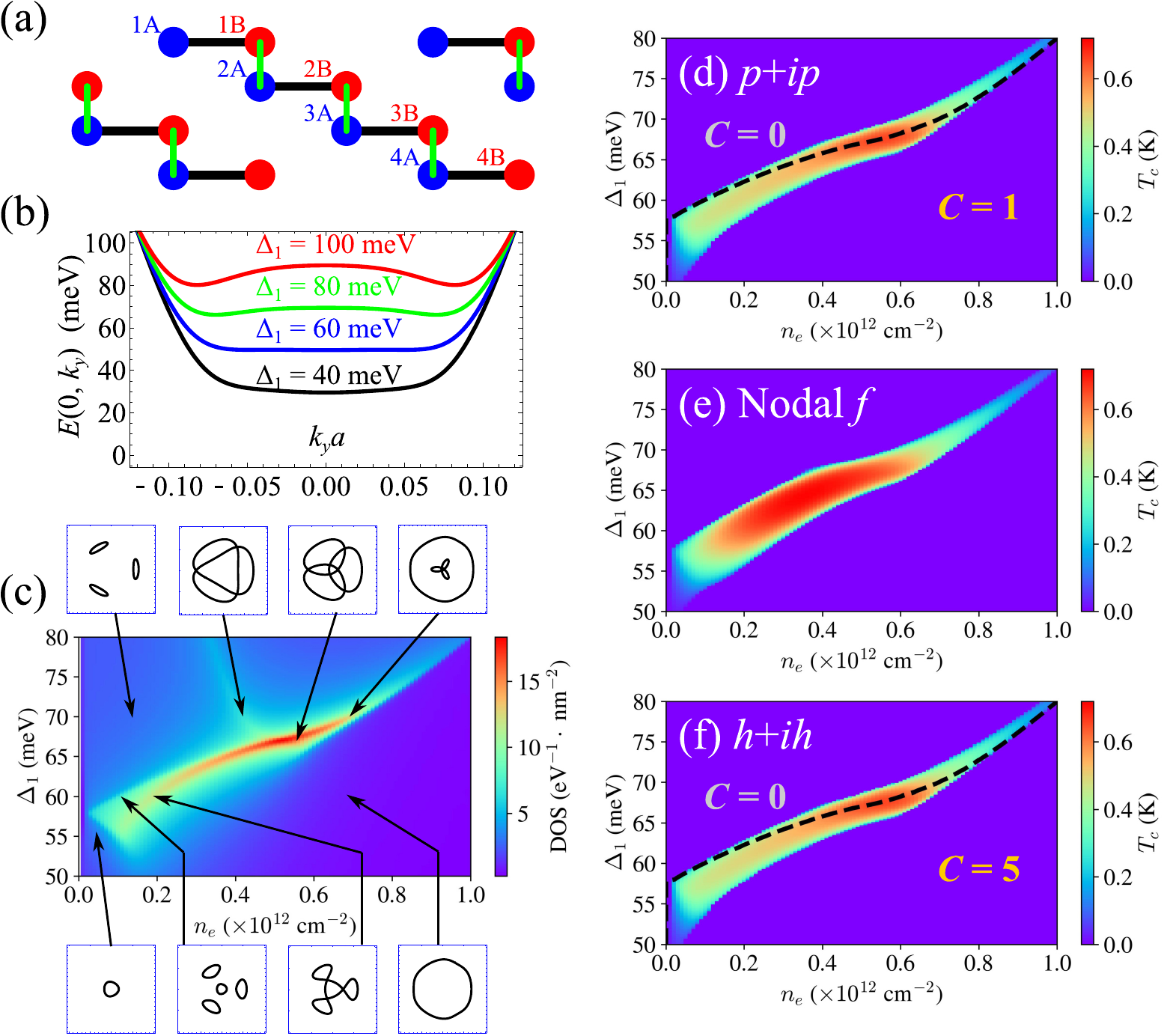}
		\caption{Setup and summary of main results. (a) The rhombohedral stacking pattern of tetralayer graphene. The blue (red) dots sublattice A (B) sites; the layer is labeled by 1, 2, 3, and 4 from top to bottom. (b) The dispersion of the first electron band (from charge neutrality) along $y$ direction relative to the valley $+K$ point with several $\Delta_1$ (electric potential, not the SC order parameter), tuned by the applied displacement field. $a$ is the graphene lattice constant. (c) Density of states (assuming spin and valley polarization) as a function of doping density ($n_e$) and $\Delta_1$. Several representative Fermi surfaces are also illustrated. (d) $T_c$ of $p+ip$ SC with $g_1=200$ meV$.$nm$^2$. (e) $T_c$ of nodal $f$ SC with $g_3=180$ meV$.$nm$^2$. (f) $T_c$ of $h+ih$ SC with $g_5=200$ meV$.$nm$^2$. The results in (d)-(f) suggest that intravalley spin-polarized SC is strongest near the VHS at $\Delta_1=67$ meV that shows six Fermi contour crossing points in the $\vex{k}$ space. (d) and (f) show the identical $T_c$'s within our numerical accuracy. The black dashed lines in (d) and (f) separate two distinct topological phases with Chern numbers $\mathcal{C}=1$ and $\mathcal{C}=0$. We have used $\approx 2\times10^6$ $k$ points and assumed spin and valley polarization in (c)-(f).
		}
		\label{Fig:Main_results}
	\end{figure}

	Superconductivity in the graphene-based materials (and moir\'e transition metal dichalcogenides) often requires intervalley pairings between two time-reversal related bands \cite{XuC2018,WuF2018,WuF2019,YouYZ2019,LianB2019a,LeeJY2019,SlagleK2020,ScheurerMS2020a,ChristosM2020,KhalafE2021,ChouYZ2021,GhazaryanA2021,CeaT2022c,ChatterjeeS2022b,SzaboAL2022,LiZ2023,Jimeno-PozoA2023,PantaleonPA2023,CrepelV2021,ChristosM2022,ZhuJ2024,KimS2024,ChristosM2024,XieF2024,GuerciD2024} because the ``Fermi surface nesting'' guarantees zero-temperature SC in the presence of an arbitrarily small attractive interaction. A recent rhombohedral tetralayer graphene experiment discovers a novel time-reversal broken SC (with $T_c\sim 300$ mK), emerging from a spin-polarized valley-polarized normal state \cite{HanT2024a}. Such a superconducting order persists a large out-of-plane magnetic field ($B_{\perp,c}>0.9$ T) and shows hysteresis of $R_{xy}$ under an out-of-plane magnetic field. The experimental evidence suggests a possible chiral topological SC \cite{VolovikGE1988,ReadN2000,HasanMZ2010,QiXL2011}, which has not been conclusively confirmed in any existing experiment. At the conceptual level, the existence of SC from intravalley pairing is also highly nontrivial. The trigonal warping in graphene multilayers generically results in non-circular Fermi surfaces \cite{ZhangF2010}, sabotaging the Fermi surface nesting for pairings.

	In this work, we study the possible SC arising from a spin-polarized valley-polarized normal state in the rhombohedral stacked tetralayer graphene [Fig.~\ref{Fig:Main_results}(a)]. First, we analyze the single-particle band structure in the electron doping [Fig.~\ref{Fig:Main_results}(b)] and reveal a multitude of Fermi surface transitions associated with Van Hove singularities (VHSs) as shown in Fig.~\ref{Fig:Main_results}(c). Then, we construct a minimal model that describes the intravalley spin-polarized pairing with an odd angular momentum, e.g., $p$-wave, $f$-wave, and $h$-wave. If the $p$-wave pairing channel dominates, the resulting SC is the $p+ip$ SC \cite{VolovikGE1988,ReadN2000,HasanMZ2010,QiXL2011,KallinC2016} (the class D in the ten-fold way classification \cite{AltlandA1997,SchnyderAP2008,HasanMZ2010}), with two distinct phases with different Chern numbers as shown in Fig.~\ref{Fig:Main_results}(d). If the $f$-wave channel dominates, the SC is a nodal topologically trivial SC. For the $h$-wave case, we find a $h+ih$ chiral SC of class D, similar to the $p$-wave case. Based on a mean-field analysis, we identify the ``magic condition'' (i.e., the optimal displacement field and doping) that favors intravalley SC without Fermi surface nesting. We also show that a finite threshold is required for realizing SC, generically resulting in narrow SC regions. We further point out that the Kohn-Luttinger mechanism \cite{KohnW1965,GhazaryanA2021,GonzalezJ2019,GuerciD2024} (i.e., the screened Coulomb interaction) may explain the intravalley SC observed in the tetralayer graphene experiment \cite{HanT2024a}, and a $p+ip$ chiral SC can be realized in this situation. Finally, we argue that the same intravalley SC may arise in other rhombohedral graphene multilayers.

    The rest of this paper is organized as follows: In Sec.~\ref{Sec:Single-particle}, we review the single-particle dispersion of the ABCA graphene. We then discuss the pairing model without specifying a particular pairing mechanism and formulate the mean-field calculations in Sec.~\ref{Sec:Pairing_model}. The possible pairing states of different pairing channels are identified in Sec.~\ref{Sec:Pairing_symmetry}, providing constraints for the intravalley spin-polarized SC. We also show the quantitative results and discuss the most favorable conditions for realizing the intravalley SC in Sec.~\ref{Sec:Condition}. In Sec.~\ref{Sec:Mechanism}, we argue that the Kohn-Luttinger mechanism is a plausible explanation for the intravalley SC based on our analysis. Several subtle issues and implications of our theory are discussed in Sec.~\ref{Sec:Discussion}. We also provide technical details in appendices. The $k\cdot p$ Hamiltonian for the ABCA graphene is given in Appendix~\ref{App:h_k}. The derivation of mean-field theory can be found in Appendix~\ref{App:MFT}. In Appendix~\ref{App:mixing}, we discuss the angular momentum mixing interaction, which is allowed by the three-fold rotational symmetry but is irrelevant to this work.

\section{Single-particle model}\label{Sec:Single-particle}
    
	We study the rhombohedral tetralayer graphene (i.e., with a chiral ABCA stacking pattern \cite{ZhangF2010}). The single-particle Hamiltonian of valley $\tau K$ and spin $s$ can be described by
	\begin{align}\label{Eq:H_0}
		\hat{H}^{(\tau,s)}_{0}=\sum_{\vex{k}}\Psi^{\dagger}_{\vex{k}}\hat{h}^{(\tau)}(\vex{k})\Psi_{\vex{k}},
	\end{align}
	where $\vex{k}=(k_x,k_y)$ is the 2D wavevector relative to the valley wavevector $\tau \vex{K}$, $\Psi$ is a 8-dimensional field incorporating layers and sublattices, and $\hat{h}^{(\tau)}(\vex{k})$ is a 8-by-8 matrix characterizing the details of band structures \cite{ZhouH2021,GhazaryanA2023,ChouYZ2022b} (see Appendix~\ref{App:h_k}). The low energy sites are the 1A (A sublattice on the top layer) and the 4B (B sublattice on the bottom layer) sites [labeled in Fig. \ref{Fig:Main_results}(a)] because the interlayer hybridization [green bonds in Fig.~\ref{Fig:Main_results}(a)] generically pushes other sites to higher energies. The band structure can be tuned by a perpendicular electric field (i.e., a displacement field), which induces imbalanced electrostatic potentials in layers, described by a single parameter $\Delta_1$ (electric potential, not the SC order parameter). With a sufficiently large $|\Delta_1|$ (which is always true in this work), the layer and sublattice are also polarized in the low-energy bands. In our convention, the energy difference between the top and bottom layers is $2|\Delta_1|$, and the low-energy electron band corresponds to the 4B sites for $\Delta_1>0$.
	
	The electron-doped band structure is sensitive to the value of $\Delta_1$. As shown in Fig.~\ref{Fig:Main_results}(b), the electron band bottom can become extremely flat for $\Delta_1\approx60$ meV. With a more careful analysis, we find that the ``flattest'' band, defined by the largest density of states (DOS), is around $\Delta_1=67$ meV, corresponding to a bifurcation point of the VHS peaks in Fig.~\ref{Fig:Main_results}(c). The corresponding Fermi surface at that point shows six crossings as illustrated in Fig.~\ref{Fig:Main_results}(c). (See Ref.~\cite{GhazaryanA2023} for similar single-particle results.) Notably, the Fermi surfaces in the same regime are not circular but typically manifest complicated Fermi surfaces [Fig.~\ref{Fig:Main_results}(c)] that reveal the absence of Fermi surface nesting for an intravalley pairing SC.  However, SC can still be realized without exact Fermi surface nesting as discussed in the pair density wave (PDW) orders \cite{AgterbergDF2020} and the recently proposed interband SC in the magic-angle twisted bilayer graphene with proximity-induced spin-orbit couplings \cite{ChouYZ2024}. Next, we construct a simplified pairing model (without specifying the pairing mechanism) and examine the possible intravalley spin-polarized SC around $\Delta_1=67$ meV in the rhombohedral tetralayer graphene.

    \section{Pairing interaction and mean-field theory}\label{Sec:Pairing_model}
    
    We are interested in pairing within the same valley and the same spin. In the presence of a large displacement field (i.e., large $|\Delta_1|$), the layer and sublattice are also polarized, leaving no internal degrees of freedom in this problem. In our convention, the low-energy electron-doped band is polarized to the 4B site. We consider the $+K$ valley and up spin without loss of generality. The intravalley spin-polarized pairing Hamiltonian (interaction in the Cooper channel) can be expressed by
	\begin{align}
		\hat{H}_I=\frac{1}{2\mathcal{A}}\sum_{\vex{k},\vex{k}'} V_{\vex{k},\vex{k}'}\psi^{\dagger}_{\vex{k}}\psi^{\dagger}_{-\vex{k}}\psi_{-\vex{k}'}\psi_{\vex{k}'},
	\end{align}
	where $\mathcal{A}$ is the 2D area, $V_{\vex{k},\vex{k}'}$ is the pairing interaction and $\psi$ is the fermionic field of 4B site with $+K$ valley and up spin. We have suppressed the layer, sublattice, and spin indexes of $\psi$ for simplicity.

    \subsection{Simplified pairing interaction}
    
	Computation with the general $V_{\vex{k},\vex{k}'}$ is challenging, especially since a very fine $\vex{k}$ mesh is required for describing the details of VHS. Thus, we consider a simplified {anisotropic} pairing interaction that depends only on the relative angle between $\vex{k}$ and $\vex{k}'$ and decompose the pairing interaction in terms of Fourier harmonics as follows:
	\begin{align}\label{Eq:g_l}
		V_{\vex{k},\vex{k}'}\approx&-\sum_{l=0}^{\infty}g_l\cos\left[l\left(\varphi_{\vex{k}}-\varphi_{\vex{k}'}\right)\right],
	\end{align}
	where $g_l$ is the attractive pairing interaction with angular momentum $l$, and $\varphi_{\vex{k}}$ is the angle of $\vex{k}$ relative to the $x$ axis. For the intravalley spin-polarized SC, only the odd angular momenta are allowed due to the antisymmetrization. The situation here is different from the intervalley pairing in the graphene-based material, in which valleys and sublattices can fulfill the antisymmetrization requirement, resulting in essentially $\vex{k}$-independent pairings \cite{WuF2019,ChouYZ2021a}. 
    
    The formal angular expansion under the three-fold rotational symmetry allows for terms that mix Cooper pairs with angular momenta differing by a multiple of three. However, these angular momentum mixing terms can be ignored in this work. The leading contribution with the angular momentum difference $\Delta l=\pm 3$ mixes singlet and triplet pairings, which are irrelevant to our situation where only triplet pairings are allowed. Moreover, the triplet-triplet mixing terms are higher-order processes (e.g., $\Delta l=\pm6,\pm 12$), which are unimportant to this work. See an extended discussion in Appendix~\ref{App:mixing}. The anisotropic pairing in Eq~(\ref{Eq:g_l}) is the main assumption in this work, which ignores the $|\vex{k}|$ and $|\vex{k}'|$ dependence on the interaction \cite{pairing_int}. As we shall discuss later, the simplified interaction allows for an efficient formulation for computing the transition temperature $T_c$ numerically.

	
    \subsection{Mean-field theory}

    The pairing potential in Eq.~(\ref{Eq:g_l}) can be factorized, using the trigonometric identity $\cos[l(\varphi_{\vex{k}}-\varphi_{\vex{k}'})]=\cos\left(l\varphi_{\vex{k}}\right)\cos\left(l\varphi_{\vex{k}'}\right)+\sin\left(l\varphi_{\vex{k}}\right)\sin\left(l\varphi_{\vex{k}'}\right)$. The pairing interaction of the angular momentum $l$ channel can be expressed by 
    \begin{align}
		\hat{H}_{I,l}=&\frac{-g_l}{2\mathcal{A}}\sum_{\vex{k},\vex{k}'} \cos\left[l\left(\varphi_{\vex{k}}-\varphi_{\vex{k}'}\right)\right]\psi^{\dagger}_{\vex{k}}\psi^{\dagger}_{-\vex{k}}\psi_{-\vex{k}'}\psi_{\vex{k}'}\\
		\label{Eq:H_Il_decomp}=&\frac{-2g_l}{\mathcal{A}}\sum_{\vex{k},\vex{k}'} '\left\{\begin{array}{r}
			[\psi^{\dagger}_{\vex{k}}\cos\left(l\varphi_{\vex{k}}\right)\psi^{\dagger}_{-\vex{k}}][\psi_{-\vex{k}'}\cos\left(l\varphi_{\vex{k}'}\right)\psi_{\vex{k}'}]\\[2mm]
			+[\psi^{\dagger}_{\vex{k}}\sin\left(l\varphi_{\vex{k}}\right)\psi^{\dagger}_{-\vex{k}}][\psi_{-\vex{k}'}\sin\left(l\varphi_{\vex{k}'}\right)\psi_{\vex{k}'}]
		\end{array}
		\right\},
	\end{align}
    where $\sum'$ denotes summation over half of the $\vex{k}$ mesh with $\vex{k}=0$ excluded. The two terms in Eq.~(\ref{Eq:H_Il_decomp}) correspond to the scattering channels $\cos(l\varphi_{\vex{k}})$ and $\sin(l\varphi_{\vex{k}})$. For $l=1$, these two terms describe the $p_x$ and $p_y$ scatterings.

	Next, we perform the Hubbard-Stratonovich decoupling for Eq.~(\ref{Eq:H_Il_decomp}) and consider the static translational invariant saddle point solution, equivalent to the standard mean-field theory. The single-band projection onto the first electron band is also employed. Finally, the fermions are integrated out, and a Landau theory is constructed by perturbing the ordering parameter. The detailed derivation is provided in Appendix~\ref{App:MFT}. We summarize the main results in the following.
	
	For a given pairing channel $l$, the Landau free energy density can be expressed by
	\begin{align}
		\nonumber\frac{\mathcal{F}}{\mathcal{A}}\approx&\left[\begin{array}{cc}
			\Delta_{l,1}^* & \Delta_{l,2}^*
		\end{array}\right]\!\left[\begin{array}{cc}
		\frac{1}{2g_l}-a_1(T) & -a_{12}(T)\\[2mm]
		-a_{12}(T) & \frac{1}{2g_l}-a_2(T)
		\end{array}\right]\!\left[\begin{array}{c}
			\Delta_{l,1}\\[2mm]
			\Delta_{l,2}
		\end{array}\right]\\
        \nonumber&+b_1(T)\left|\Delta_{l,1}\right|^4+b_2(T)\left|\Delta_{l,2}\right|^4\\
        \label{Eq:Landau_f}&+b_{12}(T)\left[\left|\Delta_{l,1}\right|^2\left|\Delta_{l,2}\right|^2+\left(\Delta_{l,1}\Delta_{l,2}^*\right)^2+\left(\Delta_{l,1}^*\Delta_{l,2}\right)^2\right],
	\end{align}
	where only the quadratic and quartic orders are kept. The expressions of $a_1$, $a_2$, $a_{12}$, $b_1$, $b_2$, and $b_{12}$ are provided in Appendix~\ref{App:MFT}. The order parameters are defined by
	\begin{subequations}\label{Eq:OP}
		\begin{align}
			\Delta_{l,1}=&-\frac{2g_l}{\mathcal{A}}\sum_{\vex{k}}'\cos\left(l\varphi_{\vex{k}}\right)\left\langle\psi_{-\vex{k}}\psi_{\vex{k}}\right\rangle,\\
			\Delta_{l,2}=&-\frac{2g_l}{\mathcal{A}}\sum_{\vex{k}}'\sin\left(l\varphi_{\vex{k}}\right)\left\langle\psi_{-\vex{k}}\psi_{\vex{k}}\right\rangle.
		\end{align}
	\end{subequations}
	
    The linearized gap equation can be derived by enforcing the vanishing quadratic term in Eq.~(\ref{Eq:Landau_f}) and is given by
	\begin{align}\label{Eq:LGE}
		\left[\begin{array}{c}
			\Delta_{l,1}\\[2mm]
			\Delta_{l,2}
		\end{array}\right]=(2g_l)\left[\begin{array}{cc}
			a_1(T_c) & a_{12}(T_c)\\[2mm]
			a_{12}(T_c) & a_2(T_c)
		\end{array}\right]\left[\begin{array}{c}
			\Delta_{l,1}\\[2mm]
			\Delta_{l,2}
		\end{array}\right].
	\end{align}
    The $T_c$ can be extracted by solving the above eigenvalue problem.
	In Eq.~(\ref{Eq:LGE}), the factor of 2 in $g_l$ is due to summing over half of the $\vex{k}$ mesh, avoiding double counting in the single-particle term. The eigenvector of Eq.~(\ref{Eq:LGE}) contains the information of the underlying pairing symmetry. In the case of doubly degenerate solutions, we need to analyze the $O(\Delta^4)$ terms in Eq.~(\ref{Eq:Landau_f}). Specifically, the sign of the $b_{12}$ determines the relative phase of $\Delta_{l,1}$ and $\Delta_{l,2}$, as we will discuss for the $p$-wave and the $h$-wave cases.
	
	The tractable expression of Eq.~(\ref{Eq:LGE}) is a consequence of the simplified pairing interaction in Eq.~(\ref{Eq:g_l}). For a general $V_{\vex{k},\vex{k}'}$, one needs to diagonalize a dense matrix with the dimension proportional to the number of $\vex{k}$ points \cite{ChouYZ2021,ChouYZ2022a,ChouYZ2022b} (around $2\times 10^6$ $\vex{k}$ points are used in this work), which is a numerically challenging task. Note that a fine $\vex{k}$-space mesh is required so that the VHS, which is crucial in our theory, can be characterized correctly.
	
	In our model, $g_l$'s are treated as free parameters. Specifically, we focus on the $p$-wave ($l=1$), $f$-wave ($l=3$), and $h$-wave ($l=5$) pairing channels. The angular momentum mixing pairing (such as $p+f$) will not be discussed in this work. We aim to investigate the conditions of realizing an intravalley spin-triplet SC (e.g., $g_l$, $\Delta_1$, and $n_e$) based on our minimal model, providing some constraints on the candidate microscopic theories. In Sec.~\ref{Sec:Mechanism}, we will discuss the screened Coulomb interaction as a potential pairing mechanism, providing estimated $g_l$'s.

	\begin{figure}[t!]
		\includegraphics[width=0.325\textwidth]{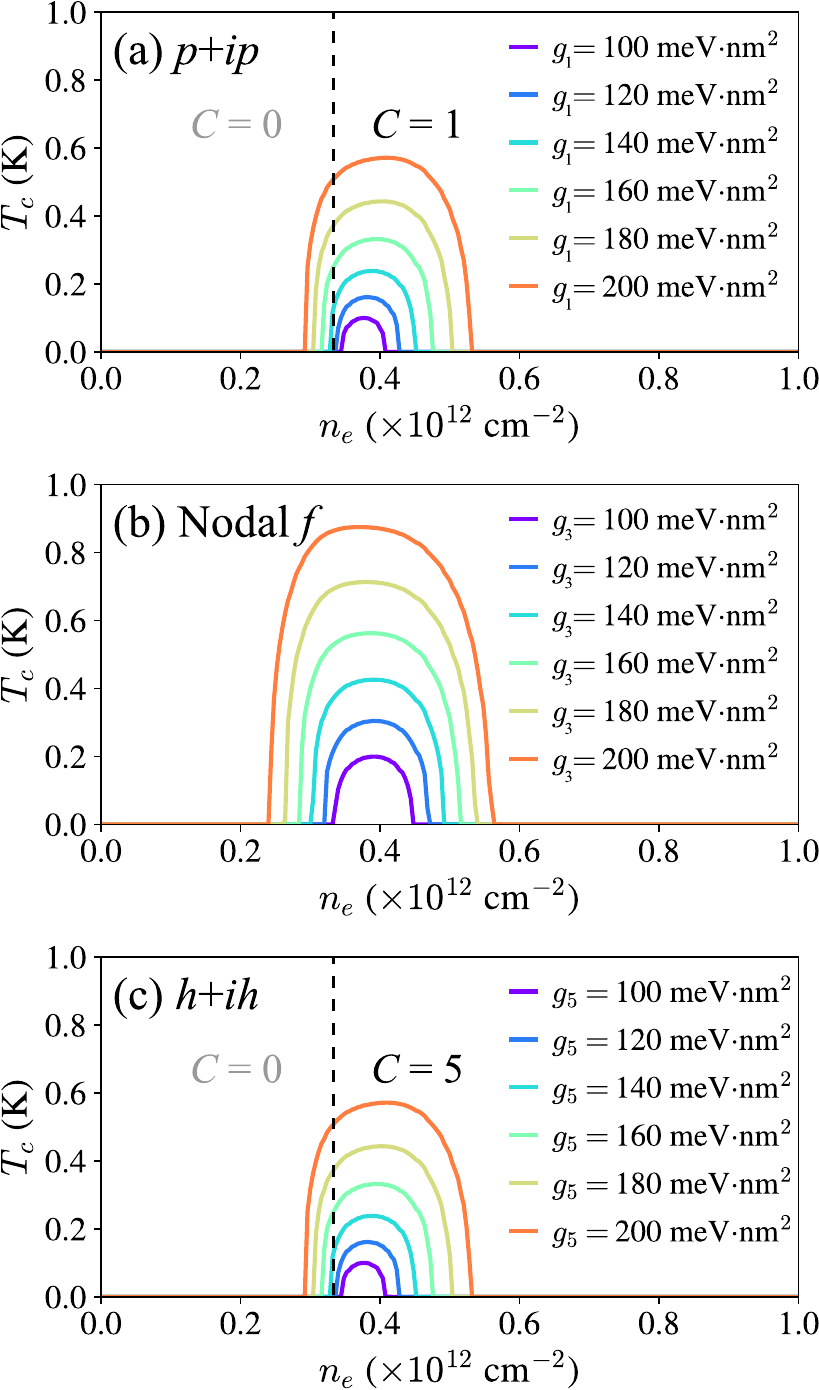}
		\caption{$T_c$ as a function of doping with $\Delta_1=65$ meV and a few representative values of $g_l$'s. (a) The $p$-wave case. (b) The $f$-wave case. (c) The $h$-wave case. For (a) and (c), there is a topological phase transition at $n_e\sim0.33\times 10^{12}$cm$^{-2}$, separating the trivial and topological phases. Note that a finite $T_c$ does not imply a finite spectral gap, and the $T=0$ gap closes at the topological phase transition. The results of $T_c$ in (a) and (c) are identical within our numerical accuracy. In all cases, we vary the value of $g_l$ and plot the corresponding $T_c$ for each $n_e$.}
		\label{Fig:Tc}
	\end{figure}
	
	\section{Pairing symmetry}\label{Sec:Pairing_symmetry}

        In this section, we focus on the symmetry and topology of the SC derived from the $p$-wave ($l=1$), $f$-wave ($l=3$), and $h$-wave ($l=5$) pairing channels. We identify chiral SC for the $p$-wave and $h$-wave cases and nodal SC for the $f$-wave case. The topological phase boundary for the chiral SC is also discussed.

        \subsection{Chiral superconductivity: $p+ip$ and $h+ih$ pairings}
    
	We first investigate the $p$-wave case ($l=1$). The two components of order parameters correspond to $p_x$ ($\Delta_{1,1}$) and $p_y$ ($\Delta_{1,2}$) orders. In the parameter regime explored in this work, we find the parameters in the free energy [Eq.~(\ref{Eq:Landau_f})] $a_1=a_2$ and $a_{12}=0$ within numerical accuracy (i.e., $\Delta_{1,1}$ and $\Delta_{1,2}$ components are degenerate), indicating that the importance of analyzing the quartic terms, the second and third lines in Eq.~(\ref{Eq:Landau_f}). We then compute $b_1$, $b_2$, and $b_{12}$, and find that the quartic terms can be expressed by $u(T)\left(|\Delta_{l,1}|^2+|\Delta_{l,2}|^2\right)^2
	+u'(T)\left(\Delta_{l,1}\Delta_{l,2}^*-\Delta_{l,2}\Delta_{l,1}^*\right)^2$ with $u(T)>0$ and $u'(T)>0$. The positivity of $u'(T)$ favors a solution with a pure imaginary $\Delta_{l,1}\Delta_{l,2}^*$. Thus, a chiral $p+ip$ SC of class D \cite{VolovikGE1988,ReadN2000,HasanMZ2010,QiXL2011,KallinC2016} is realized and there are two distinct phases: A trivial phase [Chern number $\mathcal{C}=0$] and a topologically nontrivial phase with a Chern number $\mathcal{C}=1$, manifesting chiral Majorana edge states on the sample boundary \cite{VolovikGE1988,ReadN2000,HasanMZ2010,QiXL2011,KallinC2016}. In Fig.~\ref{Fig:Main_results}(d), we plot the phase boundary (black dashed line) which is determined by $E(\vex{k}=0)$ \cite{GhoshP2010}, depending only on $\Delta_1$ (the electric potential induced by the applied displacement field) and $n_e$. The chiral $p+ip$ SC is generically gapped except for the transition line separating $\mathcal{C}=0$ and $\mathcal{C}=1$. We also note that the intravalley pairing carries finite momenta, making it a PDW with a Fulde-Ferrell-like plane-wave order parameter \cite{AgterbergDF2020}, distinct from other proposed $p+ip$ SC \cite{KallinC2016}.

	The $h$-wave case ($l=5$) also show degenerate order parameters at the quadratic order of the free energy, and the quartic order terms favor a chiral $h+ih$ SC, similar to the $p$-wave case. There are also two distinct phases: A trivial phase [Chern number $\mathcal{C}=0$] and a topologically nontrivial phase with a Chern number $\mathcal{C}=5$. In Fig.~\ref{Fig:Main_results}(f), the topological phase transition line (black dashed line) is determined by determined by $E(\vex{k}=0)$ \cite{GhoshP2010}, the same as the $p+ip$ case.

    \subsection{Nodal superconductivity: $f$-wave pairing}
	Here, we investigate the $f$-wave case corresponding to $l=3$. The $f$-wave pairing is qualitatively different from the $p$-wave and $h$-wave cases. The two components of the order parameters correspond to the weighting factors $\cos(3\varphi_{\vex{k}})$ ($\Delta_{3,1}$) and $\sin(3\varphi_{\vex{k}})$ ($\Delta_{3,2}$). We find the parameters in the free energy [Eq.~(\ref{Eq:Landau_f})] $a_1<a_2$ and $a_{12}=0$ within the numerical accuracy, suggesting that $\Delta_{3,2}$ generally dominates over the $\Delta_{3,1}$ in the parameter regime we explore. The results indicate a nodal $f$-wave SC which supports gapless excitations and carries a finite momentum because of the intravalley nature of the state. The intravalley nodal $f$-wave SC here can be viewed as a PDW \cite{AgterbergDF2020} with a Fulde-Ferrell-like plane-wave order parameter.

	\begin{figure}[t!]
		\includegraphics[width=0.325\textwidth]{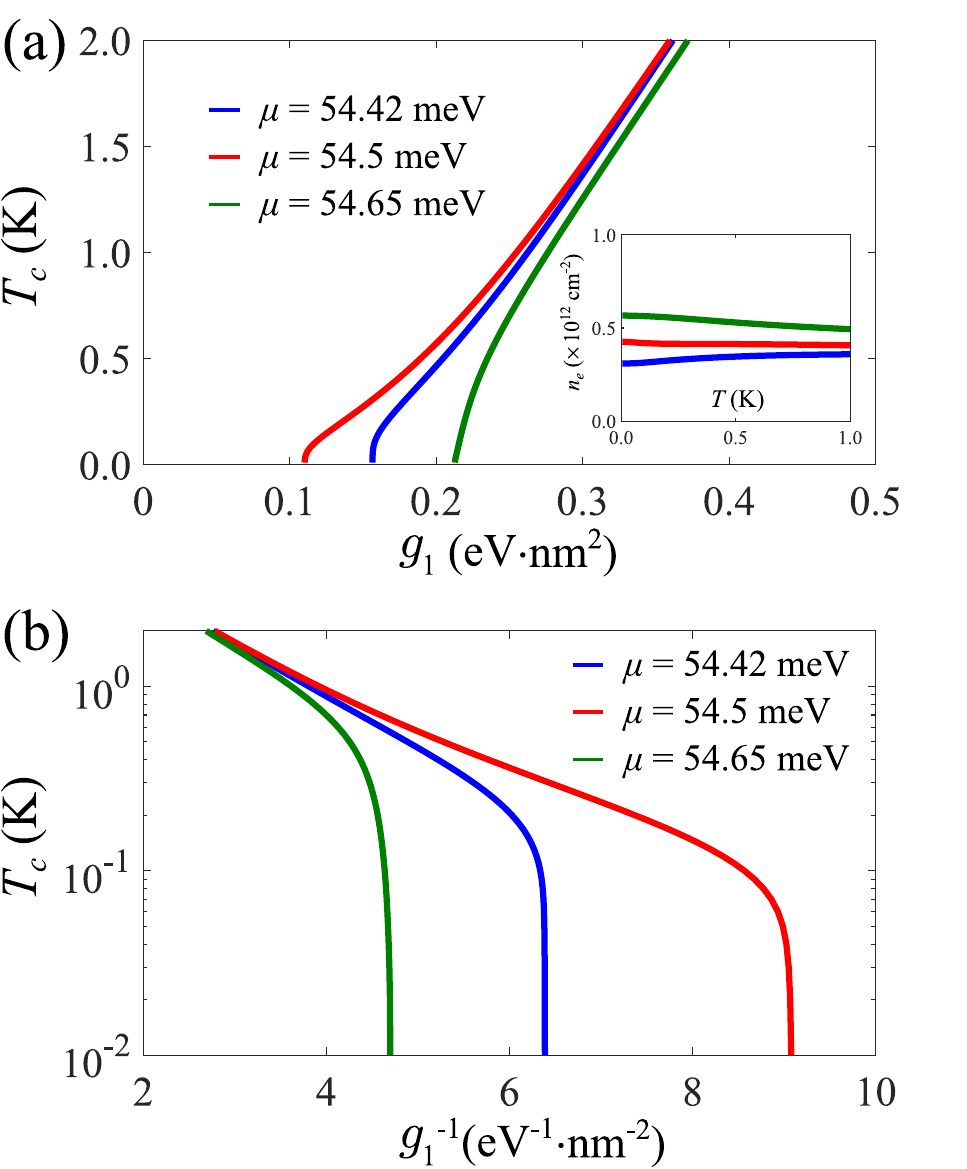}
		\caption{The $T_c$ as a function of $g_1$ for the $p$-wave chiral SC. We consider $\Delta_1=65$ meV and a few representative values of the chemical potentials $\mu$. (a) $T_c$ as a function of $g_1$. The results show a sharp termination of $T_c$, instead of the exponentially small $T_c$ in the BCS SC. Inset: The doping density ($n_e$) as a function of temperature. (b) $T_c$ as a function of $g_1^{-1}$. For each case, $T_c$ deviates from the well-known exponential dependence for a sufficiently large $g_1^{-1}$ (equivalently, a sufficiently small $g_l$). The results here clearly indicate the existence of a threshold coupling constant below which SC vanishes exactly at $T=0$. The threshold value is non-universal, depending on $\mu$, $\Delta_1$, and other single-particle model parameters. The results of $f$-wave SC ($l=3$) are similar.}
		\label{Fig:NonBCS}
	\end{figure}

    \section{Conditions for observable intravalley superconductivity}\label{Sec:Condition}

    Now, we investigate the conditions for observable intravalley SC (e.g., $T_c>0.02$ K). We numerically solve the linearized gap equation given by Eq.~(\ref{Eq:LGE}) for each pairing channel. In Figs.~\ref{Fig:Main_results}(d)-\ref{Fig:Main_results}(f), we plot $T_c$'s as functions of the doping density $n_e$ and $\Delta_1$ for the $p$-wave (with $g_1=200$ meV$\cdot$nm$^{2}$), $f$-wave (with $g_3=180$ meV$\cdot$nm$^{2}$), and $h$-wave (with $g_5=200$ meV$\cdot$nm$^{2}$) cases, respectively. The values of $g_l$'s are chosen such that the maximal $T_c$ is comparable to the experiment \cite{HanT2024a}. We find that the peak of $T_c$ follows the VHS, and the maximal $T_c$ happens close to the largest density of states location at $\Delta_1\approx 67$ meV, the bifurcation point in the VHS peaks. Remarkably, SC is absent in the left branch of the VHS for $\Delta_1>67$meV for the values of $g_l$'s used here. The resulting superconducting region is a stripe, qualitatively consistent with the experimental findings \cite{HanT2024a}. To understand the doping dependence further, we plot the $T_c$ as a function of the doping density $n_e$ at $\Delta_1=65$ meV in Fig.~\ref{Fig:Tc}, showing how $T_c$ depends on the pairing strength $g_l$. We find that the SC drops quite rapidly away from VHS, qualitatively different from the previous studies of acoustic-phonon SC \cite{ChouYZ2021,ChouYZ2022a,ChouYZ2022b}.
    

    An interesting result is that the linearized gap equation [Eq.~(\ref{Eq:LGE})] yields the same $T_c$ (for the same $g_l$) for all odd $l$ that is not divisible by 3 [as shown in Fig.~\ref{Fig:Main_results}(f) for the $p$-wave ($l=1$) and Fig.~\ref{Fig:Tc}(c) for the $h$-wave ($l=5$)]. This hidden symmetry is likely an artifact of our simplified pairing interaction in Eq.~(\ref{Eq:g_l}), which should be lifted by other perturbations.

    The intravalley SC discussed in this work has some fundamental differences from the intervalley SC that arises from time-reversal-related active bands \cite{XuC2018,WuF2018,WuF2019,YouYZ2019,LianB2019a,LeeJY2019,KhalafE2021,ChouYZ2021a,GhazaryanA2021,ZhuJ2024}. Generically, the SC emerges from normal states with asymmetric Fermi surfaces under $\vex{k}\rightarrow-\vex{k}$ [see Fig.~\ref{Fig:Main_results}(c) for several representative Fermi surface contours], suggesting that SC is absent in the weak-coupling limit. We also note that observable intravalley SC was thought to be unlikely because of the complicated normal-state Fermi surfaces, but the strong VHS in the electron-doped rhombohedral tetralayer graphene makes the unexpected intravalley SC possible.
    
    In Fig.~\ref{Fig:NonBCS}(a), we show that the linearized gap equation (with  a fixed chemical potential $\mu$) of the $p$-wave case yields a threshold value of $g_l$ below which SC is completely suppressed. The doping density $n_e$ for a fixed $\mu$ varies only slightly as lowering $T$ as shown in the inset of Fig.~\ref{Fig:NonBCS}(a). Thus, we conclude that there is a finite threshold for a fixed $n_e$ situation, also. We find similar results in other cases with $l>1$. In Fig.~\ref{Fig:NonBCS}(b), we plot $T_c$ as a function of $g_1^{-1}$ and find that the results deviate from the exponential decaying behavior at some values of $g_1^{-1}$. Note that the celebrated BCS formula is given by $T_c \propto \exp\left(-1/\lambda\right)$, where $\lambda$ is the dimensionless BCS coupling constant. The existence of a threshold for realizing SC explains the narrowness and sharp termination in the SC region as shown in Fig.~\ref{Fig:Tc}. In addition, the requirement of a threshold $g_l$ also affects the shape of the superconducting region in a nontrivial way. The stripe shape superconducting region in the experiment indicates that $g_l$ cannot be arbitrarily large because a sufficiently large $g_l$ can induce SC in the upper left branch of VHS, resulting a ``y'' shape superconducting region. Our choices of $g_l$'s give the comparable $T_c$ and qualitatively the same shape of superconducting region in the experiment.
    

    \section{Possible pairing origin: Kohn-Luttinger mechanism}\label{Sec:Mechanism}

    So far, we treat the pairing interaction phenomenologically without specifying the origin. In this section, we discuss possible pairing mechanisms for the intravalley SC in rhombohedral tetralayer graphene experiment \cite{HanT2024a}. First, the intravalley spin-polairzed SC emerges from a normal state without any internal degrees of freedom, suggesting that the conventional phonon-mediated pairings \cite{WuF2018,WuF2019,LianB2019a,ChouYZ2021} are likely irrelevant. Second, the value of the $T_c$ seems to be anticorrelated with the WSe$_2$ layer: The maximal $T_c$ is found in the device without a WSe$_2$ layer, and the device with a WSe$_2$ layer proximate to the electron band yields the weakest SC \cite{HanT2024a}. The relation between the WSe$_2$ layer and $T_c$ suggests a possible Coulomb interaction induced mechanism because the WSe$_2$ layer provides a stronger dielectric screening, and SC is the strongest without WSe$_2$.
    Among all the possible Coulomb interaction induced mechanisms, isospin fluctuation \cite{WangY2021,ChouYZ2021a,DongZ2023a,DongZ2023b,DongZ2024a} and Kohn-Luttinger \cite{KohnW1965,GhazaryanA2021,GonzalezJ2019,GuerciD2024} mechanisms are widely discussed in graphene SC \cite{PantaleonPA2023}.  
    The isospin fluctuation mechanism can be ruled out if the normal state lies deep within the quarter-metal phase, where both spin and valley are polarized. Under this condition, the Kohn-Luttinger mechanism remains the most promising explanation for the observed intravalley SC in the experiment.

    Because of the situations discussed above, we focus on the static screened Coulomb interaction in the spirit of the Kohn-Luttinger mechanism \cite{KohnW1965,GhazaryanA2021,GonzalezJ2019,GuerciD2024}.
    Our strategy is to estimate the effective interaction strength $g_l$ in Eq.~(\ref{Eq:g_l}) based on a static Random Phase Approximation (RPA) theory, and then we extract superconducting properties based on the simplified model discussed in Sec.~\ref{Sec:Pairing_model}. 

    \begin{figure*}[t]
		\includegraphics[width=0.8\textwidth]{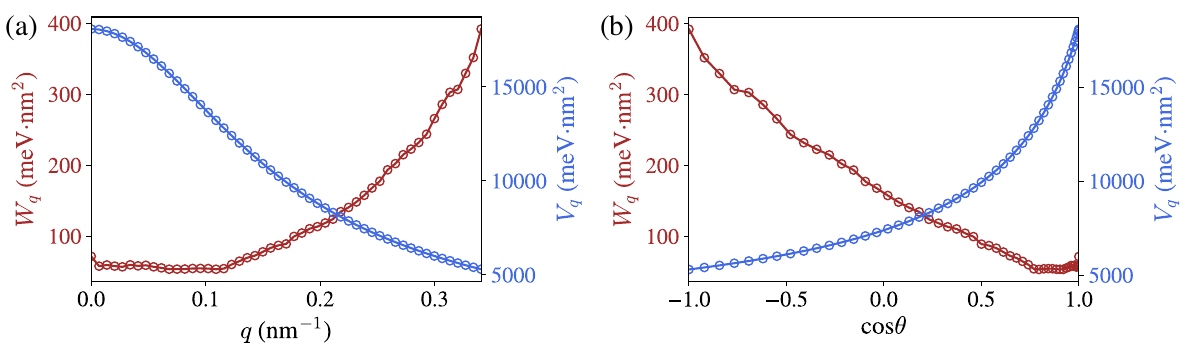}
		\caption{ The bare Coulomb potential $V_q$ in Eq.~(\ref{Eq:Vq}) and the screened Coulomb potential using Eq.~(\ref{Eq:Wq}) as a function of (a) $q$ and (b) $\cos\theta$, calculated for $\Delta_1=65$ meV, $n_e=0.4 \times 10^{12}$ cm$^{-2}$. Both $V_q$ and $W_q$ are averaged over the angle of $\mathbf{q}$ vector. After screening, the repulsive Coulomb is minimized at an intermediate $q$. In this calculation, $k_C = 0.084/a \sim 0.34$ nm$^{-1}$, and we have used a sparse $k$-grid with $\sim$ 4000 points in the calculation of polarization function.
		}
		\label{Fig:SM_Wq}
    \end{figure*}
    
    The screened Coulomb interaction, $W_{\mathbf{q}}$, in the RPA theory is given by
\begin{align}
\label{Eq:Wq}
W_{\mathbf{q}} = \frac{V_{\mathbf{q}}}{1-V_{\mathbf{q}} \chi_{0,\mathbf{q}}},
\end{align}
where
\begin{align}
\label{Eq:Vq}
V_{\mathbf{q}} = \frac{2\pi e^2}{\epsilon q} \tanh(qd_M)
\end{align}
is the bare Coulomb interaction assuming a symmetric dual metallic gates positioned at a distance $d_M$ from the sample and $\epsilon$ is the dielectric constant of the surrounding semiconductor, such as hBN. In the following calculations, we use $\epsilon = 5$ and $d_M = 10$ nm. In the long-wavelength limit, $\lim\limits_{\mathbf{q} \rightarrow 0} V_{\mathbf{q}} = 2\pi e^2 d_M/\epsilon\approx18.1$ eV$\cdot$nm$^2$, characterizing the strength of the bare Coulomb interaction.

In a spin-polarized valley-polarized state, one of the four spin-valley flavors remains metallic, while the other three are insulating. As a result, the static polarization function is dominated by particle-hole fluctuations within the metallic flavor, which corresponds to the $+K$ valley with spin-up electrons (as assumed in this work). Since the interband transitions have a minor impact on the metallic state, we project the interaction to the first conduction band (i.e., the active band). The resulting static polarization function is given by
\begin{align}
\chi_{0,\mathbf{q}} = \frac{1}{\mathcal{A}} \sum\limits_{\mathbf{k}} \frac{n_{\rm F}(E(\mathbf{k})-\mu) - n_{\rm F}(E(\mathbf{k}+\mathbf{q})-\mu)}{E(\mathbf{k}) - E(\mathbf{k}+\mathbf{q})} |\tilde\Lambda(\mathbf{k;\mathbf{q}})|^2.
\end{align}
Here, $n_{\rm F}$ is the Fermi-Dirac distribution, $\mu$ is the chemical potential, $E(\mathbf{k})$ is the energy of the active band, and $\tilde\Lambda(\mathbf{k;\mathbf{q}}) = \sum\limits_{\alpha} \phi_{\alpha}(\mathbf{k}) \phi^*_{\alpha}(\mathbf{k}+\mathbf{q})$ is the overlap between wavefunctions of the active band ($\phi_{\alpha}$'s), which sums over layers and sublattices.

To estimate the effective interaction strength $g_l$ in Eq.~(\ref{Eq:g_l}), we approximate
the screened Coulomb interaction $W_{\mathbf{q}}$ to a function of the relative angle $\theta=\varphi_{\mathbf{k}'} - \varphi_{\mathbf{k}}$ between the momenta of incoming ($\mathbf{k}'$) and outgoing ($\mathbf{k}$) scattering states. Specifically, we project $W_{\mathbf{q}}$ onto a circle with radius $k_C$ by setting $|\mathbf{k}|=|\mathbf{k}'|=k_C$. We obtain the following relations
\begin{align}
q &= 2k_C \sin \theta/2, \\
\theta &= 2 \arcsin \Big(\frac{q}{2k_C} \Big).
\end{align}
After projecting $W_{\mathbf{q}}$ onto a circle with radius $k_C$, averaged over the direction of $\mathbf{q}$ (equivalent to averaging over $\varphi_{\vex{k}'}$), $W_{\mathbf{q}}\equiv W(q)\rightarrow  W(\cos\theta)$ can be expanded in terms of $\cos(l\theta)$,
\begin{align}
W(\cos\theta) = \frac{1}{2}W_0 + \sum\limits_{l=1}^{\infty} W_l \cos(l\theta),
\end{align}
and the effective interaction in the $l$-th angular momentum channel is
\begin{align}
W_l = \frac{2}{\pi}\int_{0}^{\pi} W(\cos\theta) \cos(l\theta) d\theta,
\end{align}
which relates to the effective attraction in Eq.~(\ref{Eq:g_l}) by $g_l=-W_l$ for $l\ge 1$ and $g_0=-W_0/2$. The analysis can be extended to the cases with $|\mathbf{k}|\neq|\mathbf{k}'|$.

In Fig.~\ref{Fig:SM_Wq}, we plot the bare Coulomb potential $V_{\vex{q}}$ in Eq.~(\ref{Eq:Vq}) and the screened Coulomb potential $W_{\vex{q}}$ using Eq.~(\ref{Eq:Wq}) as functions of $q$ and $\cos\theta$, calculated for $\Delta_1=65$ meV, $n_e=0.4 \times 10^{12}$ cm$^{-2}$. Both $V_{\vex{q}}$ and $W_{\vex{q}}$ are averaged over the angle of $\mathbf{q}$ vector.
After screening, the repulsive Coulomb interaction is minimized at an intermediate value of $q$. 
For the case of $\Delta_1 = 65$ meV and $n_e=0.4 \times 10^{12}$ cm$^{-2}$, we summarize the leading interaction terms in Table \ref{tab:W_l}. 
\begin{table*}[t]
\resizebox{0.4\textwidth}{!}{
\begin{tabular}{ |c|c|c|c|c|c| }
\hline
$k_Ca$ & $0.042$ & $0.063$ & $0.084$ & $0.105$ & $0.126$ \\
\hline
$W_0$ & $200.85$ & $268.11$ & $373.01$ & $721.26$ & $1300.61$ \\
\hline
$W_1$ & $-37.64$ & $-91.82$ & $-161.42$ & $-398.67$ & $-759.22$ \\
\hline
$W_2$ & $13.49$ & $16.62$ & $23.29$ & $100.03$ & $137.19$ \\
\hline
$W_3$ & $6.13$ & $10.86$ & $8.45$ & $12.38$ & $66.09$ \\
\hline
$W_4$ & $-5.67$ & $-2.59$ & $-4.70$ & $-38.42$ & $-64.90$ \\
\hline
$W_5$ & $3.53$ & $8.03$ & $11.25$ & $46.32$ & $42.33$ \\
\hline
\end{tabular}}
\caption{
\label{tab:W_l}
Effective interactions $W_l$ in unit of meV$\cdot$nm$^2$ for the case of $\Delta_1 = 65$ meV and $n_e=0.4 \times 10^{12}$ cm$^{-2}$. Considered $k_C$ are near the Fermi momentum. $a=0.246$ nm is graphene lattice constant. In this calculation, we have used a sparse $k$-grid with $
\sim$4000 $k$ points.}
\end{table*}
In Table~\ref{tab:W_l}, the dominant effective attraction is in the $p$-wave channel, and the $f$-wave and $h$-wave channels are repulsive. In Appendix~\ref{App:mixing}, we discuss the angular momentum mixing terms allowed by the three-fold rotational symmetry and point out that these terms are irrelevant to our work. Therefore, we focus only on the $p$-wave pairing in the rest of the discussion.

To extract the value of $g_1$, we select the value of $k_C$ that is comparable to the Fermi wavevector. Based on Table~\ref{tab:W_l}, this estimate gives a $g_1\sim$ 100 to 150 meV$.$nm$^2$, corresponding a maximal $T_c\sim$ 0.1 to 0.3 K [See Fig.~\ref{Fig:Tc}(a)]. According to our analysis here and the results in Sec.~\ref{Sec:Pairing_symmetry}, the screened Coulomb interaction can give rise to observable chiral $p+ip$ SC with $T_c$ comparable to the experiment \cite{HanT2024a}. It should be noted that the analysis here is based on a sparse $k$ grid with 4000 $k$ points. A denser $k$ grid or a more accurate determination of the Fermi momentum would likely improve the precision of the calculated effective interactions. To estimate the $g_1$ more accurately, we also need to average over different $k_C$ and relax the condition $|\vex{k}|=|\vex{k}'|$. A systematic investigation of the Kohn-Luttinger mechanism should provide a better quantitative estimate of the $T_c$, which we defer for future studies.

Since other obvious candidate mechanisms (discussed in the beginning of this section) are unlikely, we speculate that the Kohn-Luttinger-like mechanism is responsible for the intravalley spin-polarized SC in rhombohedral tetralayer graphene. The main thrust and findings in the previous sections are independent of the microscopic mechanism for intravalley spin-polarized SC, and a future systematic analysis incorporating the $|\vex{k}|$-dependent pairing interaction and vertex corrections deserves a separate detailed investigation in order to decisively establish whether a Kohn-Luttinger mechanism arising from the screened Coulomb interaction (as speculated here) is indeed operational here.

    \section{Discussion}\label{Sec:Discussion}

	Now, we discuss several technical issues. In this work, we assume spin- and valley-polarized quarter metal as the normal state and adopt the single-particle band structure. Incorporating the band renormalization, the band structure acquires corrections, and the precise location and shape of SC region can be modified. Another important issue is that the microscopic interaction might not be the same as Eq.~(\ref{Eq:g_l}), and some $|\vex{k}|$ dependence is anticipated. Our model interaction can be viewed as an averaged version of the microscopic interaction, and the qualitative results should remain. For example, we expect that intravalley SC is associated with VHS, and the VHS-assisted pairing \cite{HsuYT2017,LinYP2019,LinYP2020a,ChichinadzeDV2020,WuZ2023,OjajarviR2024} might be the key for intravalley SC, regardless of the details in the band structure or the precise form of pairing interaction. A potential complication is that the vertex corrections and screening near VHS may affect the quantitative results. Additionally, competing orders, such as density-wave orders and localized states, can preempt SC. The absence of SC in the very low density limit is presumably due to the competing orders.

    Finally, we discuss the conditions for realizing the intravalley spin-polarized SC in other rhombohedral graphene systems. Based on our analysis, the intravalley SC can emerge as long as VHS appears in a quarter-metal regime, suggesting that this interesting phenomenon is \textit{not} particular to the rhombohedral tetralayer graphene. The absence of intravalley SC in Bernal bilayer and rhombohedral trilayer graphene can be understood by the experimental isospin phase diagrams, which show no sign of VHS in any of the quarter-metal regimes. The situation of rhombohedral pentalayer graphene is unclear due to lack of experimental characterization. To gain more insight, we repeat our analysis in Sec.~\ref{Sec:Mechanism} for the pentalayer graphene, assuming spin and valley polarization. We find that the VHS structure is similar to ABCA tetralayer graphene, and the screened Coulomb interaction generates a $p$-wave pairing with a $T_c$ of the order of 0.1K. The existence of intravalley SC in the rhombohedral pentalayer graphene depends on whether the VHS can be found in the quarter-metal regime. We further speculate that the same SC may also emerge in the rhombohedral hexalayer graphene, also. Our work implies the possible existence of similar intravalley SC in the rhombohedral pentalayer and rhombohedral hexalayer graphene systems, which should motivate future studies.

    Note added: During the review of this work, we learn that intravalley SC has been observed in rhombohedral pentalayer graphene \cite{HanT2024a}. The maximal $T_c$ and the corresponding superconducting regions are similar to the rhombohedral tetralayer graphene. Our analysis also suggests similar results for the rhombohedral pentalayer graphene, qualitatively consistent with the experiment.

	\begin{acknowledgments}
		We thank Long Ju and Jay D. Sau for valuable discussions. This work is supported by the Laboratory for Physical Sciences and in part by National Science Foundation to the Kavli Institute for Theoretical Physics (KITP).
	\end{acknowledgments}
	

		\appendix

    \begin{widetext}
        
    \section{Single-particle model}\label{App:h_k}
    
	 The single-particle model $\hat{h}^{(\tau)}$ is expressed by
	\begin{align}
		\label{Eq:h_kp}\hat{h}^{(\tau)}=\left[
		\begin{array}{cccccccc}
			\delta-\Delta_1 & v_0 \Pi_{\vex{k}}^{\dagger} & v_4 \Pi_{\vex{k}}^{\dagger} & v_3 \Pi_{\vex{k}} & 0 & \frac{1}{2}\gamma_2 & 0 & 0 \\
			v_0 \Pi_{\vex{k}} & -\Delta_1 & \gamma_1 & v_4 \Pi_{\vex{k}}^{\dagger} & 0 & 0 & 0 & 0 \\
			v_4 \Pi_{\vex{k}} & \gamma_1 & -\frac{\Delta_1}{3} & v_0 \Pi_{\vex{k}}^{\dagger} & v_4 \Pi_{\vex{k}}^{\dagger} & v_3 \Pi_{\vex{k}} & 0 & \frac{1}{2}\gamma_2 \\
			v_3 \Pi_{\vex{k}}^{\dagger} & v_4 \Pi_{\vex{k}} & v_0 \Pi_{\vex{k}} & -\frac{\Delta_1}{3} & \gamma_1 & v_4 \Pi_{\vex{k}}^{\dagger} & 0 & 0 \\
			0 & 0 & v_4 \Pi_{\vex{k}} & \gamma_1 & \frac{\Delta_1}{3} & v_0 \Pi_{\vex{k}}^{\dagger} & v_4 \Pi_{\vex{k}}^{\dagger} & v_3 \Pi_{\vex{k}} \\
			\frac{1}{2}\gamma_2 & 0 & v_3 \Pi_{\vex{k}}^{\dagger} & v_4 \Pi_{\vex{k}} & v_0 \Pi_{\vex{k}} & \frac{\Delta_1}{3} & \gamma_1 & v_4 \Pi_{\vex{k}}^{\dagger} \\
			0 & 0 & 0 & 0 & v_4 \Pi_{\vex{k}} & \gamma_1 & \Delta_1 & v_0 \Pi_{\vex{k}}^{\dagger} \\
			0 & 0 & \frac{1}{2}\gamma_2 & 0 & v_3 \Pi_{\vex{k}}^{\dagger} & v_4 \Pi_{\vex{k}} & v_0 \Pi_{\vex{k}} & \delta+\Delta_1 \\
		\end{array}
		\right],
	\end{align}
	where $\Pi_{\vex{k}}=\tau k_x+ik_y$ ($\tau= 1,-1$ for valleys $+K$ and $-K$ respectively), $v_j=\frac{\sqrt{3}}{2}\gamma_ja_0$, $\gamma_j$ is the bare hopping matrix element, and $a_0=0.246$ nm is the lattice constant of graphene. $\gamma_0=3.1$ eV, $\gamma_1=0.38$ eV, $\gamma_2=-0.015$ eV, $\gamma_3=-0.29$ eV, $\gamma_4=-0.141$ eV, and $\delta=-0.0105$ eV. The value of $\Delta_1$ corresponds to the out-of-plane displacement field. The basis of the matrix is (1A,1B,2A,2B,3A,3B,4A,4B) with the number representing the layer and A/B representing the sublattice. The band parameters here are obtained from Ref.~\cite{ZhouH2021} for rhombohedral trilayer graphene. The model here is equivalent to Ref.~\cite{GhazaryanA2023} and Ref.~\cite{ChouYZ2022b} with $\Delta_2$ and $\Delta_3$ set to zero.

	\section{Derivation of Mean-field theory}\label{App:MFT}

	\begin{figure}[t!]
	\includegraphics[width=0.7\textwidth]{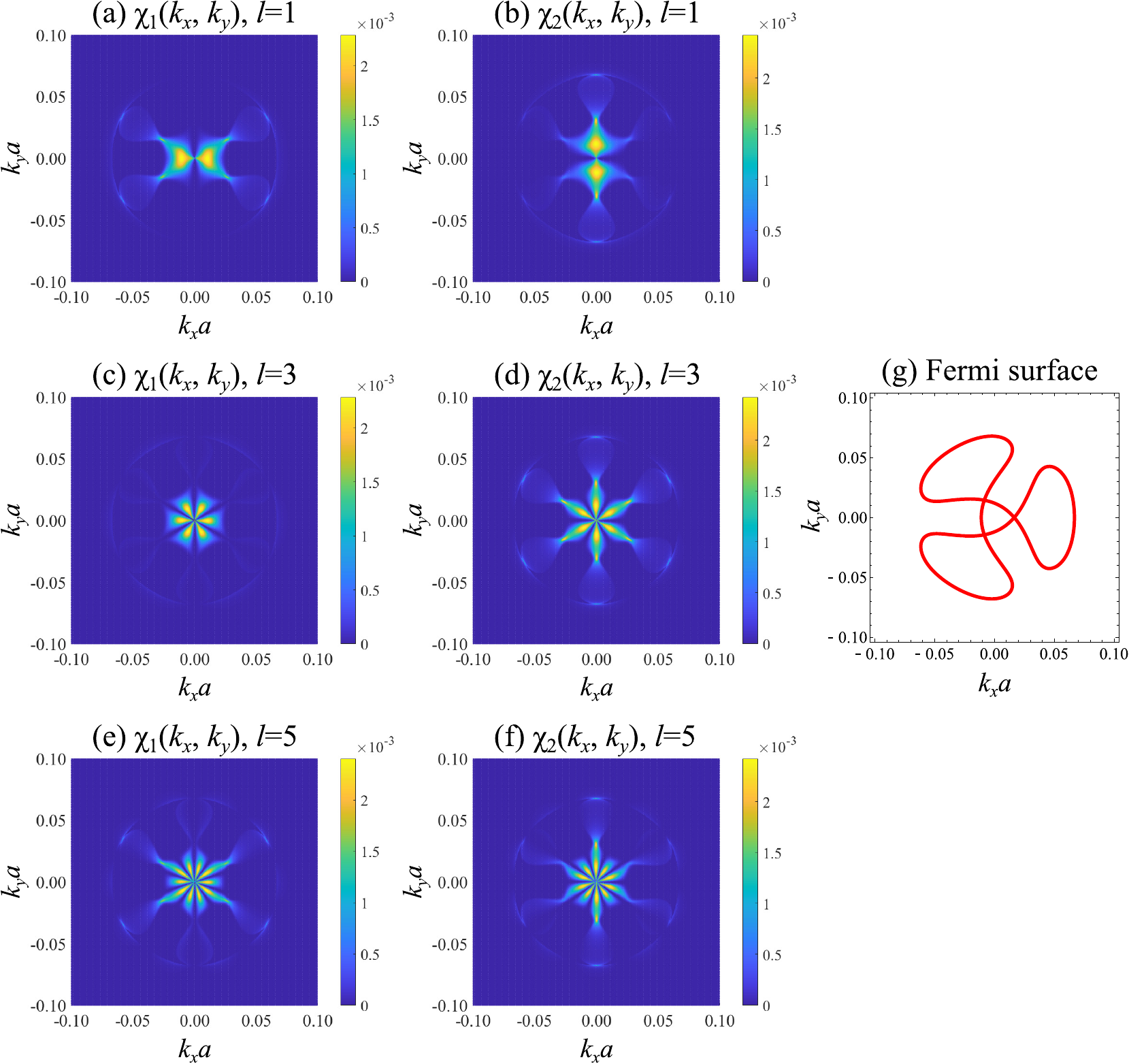}
	\caption{$\chi_1(\vex{k})$ and $\chi_2(\vex{k})$ for the $p$-wave ($l=1$), $f$-wave ($l=3$), and $h$-wave ($l=5$) cases. We consider $\Delta_1=65$ meV, $\mu=54.474$ meV, and $T=0.1$ K. (a), (b): The $p$-wave case. (c), (d): The $f$-wave case. (e), (f): The $h$-wave case. The $\chi_1(\vex{k})$'s as functions of $\vex{k}$ are plotted in
		(a), (c), and (e); The $\chi_2(\vex{k})$'s as functions of $\vex{k}$ are plotted in (b), (d), and (f). Only half of the $\vex{k}$ points are computed. We extend the results by using $\chi_{a}(-\vex{k})=\chi_{a}(\vex{k})$ with $a=1,2$. (g) The Fermi surface corresponds to $\Delta_1=65$ meV and $\mu=54.474$ meV.
		}
	\label{Fig:chi_k}
\end{figure}

	In this appendix, we formulate the mean-field theory and derive the linearized gap equation. In the imaginary-time path integral, we can decouple the pairing interaction $\hat{H}_{I,l}$ [Eq.~(\ref{Eq:H_Il_decomp})] as follows:
	\begin{align}
		\mathcal{S}_{I,l}=&\frac{-2g_l}{\mathcal{A}}\int\limits_{\tau}\sum_{\vex{k},\vex{k}}'\left\{\begin{array}{r}
			\left[\bar\psi_{\vex{k}}\cos\left(l\varphi_{\vex{k}}\right)\bar\psi_-\vex{k}\right]\left[\psi_{-\vex{k}'}\cos\left(l\varphi_{\vex{k}'}\right)\psi_{\vex{k}'}\right]\\[2mm]
			+\left[\bar\psi_{\vex{k}}\sin\left(l\varphi_{\vex{k}}\right)\bar\psi_-\vex{k}\right]\left[\psi_{-\vex{k}'}\sin\left(l\varphi_{\vex{k}'}\right)\psi_{\vex{k}'}\right]
		\end{array}
		\right\}_{\tau}\\
		\label{Eq:S_I_HB}\rightarrow&\int\limits_{\tau}\sum_{\vex{k}}'\left\{\begin{array}{c}
			\left[\Delta_{l,1}^*\cos\left(l\varphi_{\vex{k}}\right)+\Delta_{l,2}^*\sin\left(l\varphi_{\vex{k}}\right)\right]\psi(-\vex{k})\psi(\vex{k})\\[2mm]
			+\left[\Delta_{l,1}\cos\left(l\varphi_{\vex{k}}\right)+\Delta_{l,2}\sin\left(l\varphi_{\vex{k}}\right)\right]\bar\psi(\vex{k})\bar\psi(-\vex{k})
		\end{array}
		\right\}_{\tau}+\frac{\mathcal{A}}{2g_l}\int\limits_{\tau}\left(\left|\Delta_{l,1}(\tau)\right|^2+\left|\Delta_{l,2}(\tau)\right|^2\right),
	\end{align}
	where $\Delta_{l,1}$ and $\Delta_{l,2}$ are the Hubbard-Stratonovich fields. The saddle point equation gives
	\begin{align}
		\Delta_{l,1}(\tau)=-\frac{2g_l}{\mathcal{A}}\sum_{\vex{k}}'\cos\left(l\varphi_{\vex{k}}\right)\left\langle\psi_{-\vex{k}}\psi_{\vex{k}}\right\rangle(\tau),\,\,\Delta_{l,2}(\tau)=-\frac{2g_l}{\mathcal{A}}\sum_{\vex{k}}'\sin\left(l\varphi_{\vex{k}}\right)\left\langle\psi_{-\vex{k}}\psi_{\vex{k}}\right\rangle(\tau).
	\end{align}
	
	We focus on the $b$th band (the active electron band) and perform projection. The action is described by
	\begin{align}
		\nonumber\mathcal{S}=\mathcal{S}_0+\mathcal{S}_{I,l}=&\int\limits_{\tau}\sum_{\vex{k}}\bar{c}_b(\tau,\vex{k})\left[\partial_{\tau}+E(\vex{k})-\mu\right]c_b(\tau,\vex{k})\\
		\nonumber&+\int\limits_{\tau}\sum_{\vex{k}}'\left\{\begin{array}{c}
			\left[\Delta_{l,1}^*(\tau)\cos\left(l\varphi_{\vex{k}}\right)+\Delta_{l,2}^*(\tau)\sin\left(l\varphi_{\vex{k}}\right)\right]\Lambda_{b,\vex{k}}c_b(\tau,-\vex{k})c_b(\tau,\vex{k})\\[2mm]
			+\left[\Delta_{l,1}(\tau)\cos\left(l\varphi_{\vex{k}}\right)+\Delta_{l,2}(\tau)\sin\left(l\varphi_{\vex{k}}\right)\right]\Lambda_{b,\vex{k}}^*\bar{c}_b(\tau,\vex{k})\bar{c}_b(\tau,-\vex{k})
		\end{array}
		\right\}\\
		&+\frac{\mathcal{A}}{2g_l}\sum_{l=1,3}\int\limits_{\tau}\left(\left|\Delta_{l,1}(\tau)\right|^2+\left|\Delta_{l,2}(\tau)\right|^2\right),
	\end{align}
	where $c_b$ is the fermionic field of the $b$th band (which is different from $\psi$), $E(\vex{k})$ is the energy of the $b$th band, $\mu$ is the chemical potential, $\Lambda_{b,\vex{k}}=\phi^{(b)}_{\tau,n,\sigma,s}(-\vex{k})\phi^{(b)}_{\tau,n,\sigma,s}(\vex{k})$, and  $\phi^{(b)}_{\tau,n,\sigma,s}(\vex{k})$ is the wavefunction of the $b$th band with valley $\tau$, layer $n$, sublattice $\sigma$, and spin $s$. The appearance of the form factor $\Lambda_{b,\vex{k}}$ is due to the band projection.

	To derive the mean-field equations, we first assume static order parameters, i.e., $\Delta_{l,1}(\tau)=\Delta_{l,1}$ and $\Delta_{l,2}(\tau)=\Delta_{l,2}$. The imaginary-time action becomes
	\begin{align}
		\nonumber\mathcal{S}=&\frac{1}{\beta}\sum_{\omega_n}\!\sum_{\vex{k}}'\!\!\left[\begin{array}{cc}
			\bar{c}(k) & c(-k)
		\end{array}\right]\!
		\left[\begin{array}{cc}
			-i\omega_n+E(\vex{k})-\mu& \left[\Delta_{l,1}\cos\left(l\varphi_{\vex{k}}\right)\!+\!\Delta_{l,2}\sin\left(l\varphi_{\vex{k}}\right)\right]\!\Lambda_{b,\vex{k}}^*\\
			\left[\Delta_{l,1}^*\cos\left(l\varphi_{\vex{k}}\right)\!+\!\Delta_{l,2}^*\sin\left(l\varphi_{\vex{k}}\right)\right]\!\Lambda_{b,\vex{k}} & -i\omega_n-E(-\vex{k})+\mu
		\end{array}\right]\!\left[\begin{array}{c}
			c(k)\\[2mm]
			\bar{c}(-k)
		\end{array}\right]\\
		&+\frac{\mathcal{A}\beta}{2g_l}\left(\left|\Delta_{l,1}\right|^2+\left|\Delta_{l,2}\right|^2\right).
	\end{align}

	To proceed, we integrate out the fermionic fields and derive the effective action for the order parameters. The free energy density is as follows:
	\begin{align}
		\label{SEq:free_E}\mathcal{F}/\mathcal{A}=&-\frac{1}{\beta\mathcal{A}}\sum_{\omega_n}\sum_{\vex{k}}'\ln\left\{\begin{array}{c}
			\left(-i\omega_n+E(\vex{k})-\mu\right)\left(-i\omega_n-E(-\vex{k})+\mu\right)\\[2mm]
			-\left|\left[\Delta_{l,1}\cos\left(l\varphi_{\vex{k}}\right)\!+\!\Delta_{l,2}\sin\left(l\varphi_{\vex{k}}\right)\right]\!\Lambda_{b,\vex{k}}^*\right|^2
		\end{array}\right\}+\frac{1}{2g_l}\left(\left|\Delta_{l,1}\right|^2+\left|\Delta_{l,2}\right|^2\right)\\
		\approx&\frac{-1}{\beta\mathcal{A}}\sum_{\omega_n}\sum_{\vex{k}}'
		\left\{\begin{array}{c}
			-\frac{\left|\left[\Delta_{l,1}\cos\left(l\varphi_{\vex{k}}\right)\!+\!\Delta_{l,2}\sin\left(l\varphi_{\vex{k}}\right)\right]\!\Lambda_{b,\vex{k}}^*\right|^2}{\left(-i\omega_n+E(\vex{k})-\mu\right)\left(-i\omega_n-E(-\vex{k})+\mu\right)}\\[3mm]
			-\frac{1}{2}\frac{\left|\left[\Delta_{l,1}\cos\left(l\varphi_{\vex{k}}\right)\!+\!\Delta_{l,2}\sin\left(l\varphi_{\vex{k}}\right)\right]\!\Lambda_{b,\vex{k}}^*\right|^4}{\left(-i\omega_n+E(\vex{k})-\mu\right)^2\left(-i\omega_n-E(-\vex{k})+\mu\right)^2}
		\end{array}\right\}
		+\frac{1}{2g_l}\left(\left|\Delta_{l,1}\right|^2+\left|\Delta_{l,2}\right|^2\right)\\
		\nonumber=&\left[\begin{array}{cc}
			\Delta_{l,1}^* & \Delta_{l,2}^*
		\end{array}\right]\left\{\left[\begin{array}{cc}
			\frac{1}{2g_l} & 0\\[2mm]
			0 & \frac{1}{2g_l}
		\end{array}\right]-\left[\begin{array}{cc}
			a_1(T) & a_{12}(T)\\[2mm]
			a_{12}(T) & a_2(T)
		\end{array}\right]\right\}\left[\begin{array}{c}
			\Delta_{l,1}\\[2mm]
			\Delta_{l,2}
		\end{array}\right]\\
		&+b_1(T)\left|\Delta_{l,1}\right|^4+b_2(T)\left|\Delta_{l,2}\right|^4+b_{12}(T)\left[\left|\Delta_{l,1}\right|^2\left|\Delta_{l,2}\right|^2+\left(\Delta_{l,1}\Delta_{l,2}^*\right)^2+\left(\Delta_{l,1}^*\Delta_{l,2}\right)^2\right],
	\end{align}
	where
	\begin{subequations}\label{Eq:a1_a2_a12}
		\begin{align}
			a_1(T)=&-\frac{1}{\mathcal{A}}\sum_{\vex{k}}'\left[\frac{n_F(E(\vex{k})-\mu)-n_F(-E(-\vex{k})+\mu)}{E(\vex{k})+E(-\vex{k})-2\mu}\cos^2\left(l\varphi_{\vex{k}}\right)\left|\Lambda_{b,\vex{k}}\right|^2\right],\\
			a_2(T)=&-\frac{1}{\mathcal{A}}\sum_{\vex{k}}'\left[\frac{n_F(E(\vex{k})-\mu)-n_F(-E(-\vex{k})+\mu)}{E(\vex{k})+E(-\vex{k})-2\mu}\sin^2\left(l\varphi_{\vex{k}}\right)\left|\Lambda_{b,\vex{k}}\right|^2\right],\\
			a_{12}(T)=&-\frac{1}{\mathcal{A}}\sum_{\vex{k}}'\left[\frac{n_F(E(\vex{k})-\mu)-n_F(-E(-\vex{k})+\mu)}{E(\vex{k})+E(-\vex{k})-2\mu}\cos\left(l\varphi_{\vex{k}}\right)\sin\left(l\varphi_{\vex{k}}\right)\left|\Lambda_{b,\vex{k}}\right|^2\right],\\
			b_1(T)=&\frac{1}{\beta\mathcal{A}}\sum_{\vex{k}}'\sum_{\omega_n}\frac{1}{2}\frac{\left|\Lambda_{b,\vex{k}}^*\right|^4\cos^4\left(l\varphi_{\vex{k}}\right)}{\left(-i\omega_n+E(\vex{k})-\mu\right)^2\left(-i\omega_n-E(-\vex{k})+\mu\right)^2}\\
			b_2(T)=&\frac{1}{\beta\mathcal{A}}\sum_{\vex{k}}'\sum_{\omega_n}\frac{1}{2}\frac{\left|\Lambda_{b,\vex{k}}^*\right|^4\sin^4\left(l\varphi_{\vex{k}}\right)}{\left(-i\omega_n+E(\vex{k})-\mu\right)^2\left(-i\omega_n-E(-\vex{k})+\mu\right)^2}\\
			b_{12}(T)=&\frac{1}{\beta\mathcal{A}}\sum_{\vex{k}}'\sum_{\omega_n}\frac{1}{2}\frac{\left|\Lambda_{b,\vex{k}}^*\right|^4\cos^2\left(l\varphi_{\vex{k}}\right)\sin^2\left(l\varphi_{\vex{k}}\right)}{\left(-i\omega_n+E(\vex{k})-\mu\right)^2\left(-i\omega_n-E(-\vex{k})+\mu\right)^2},
		\end{align}
	\end{subequations}
	and $n_F(x)$ is the Fermi-Dirac distribution function. To simplify the expression of $b_1$, $b_2$, and $b_{12}$, we carry out the Matsubara summation
	\begin{align}
		\nonumber&\sum_{\omega_n}\frac{1}{2}\frac{1}{\left(-i\omega_n+E(\vex{k})-\mu\right)^2\left(-i\omega_n-E(-\vex{k})+\mu\right)^2}\\
		&=-\frac{n_F(E(\vex{k})-\mu)-n_F(-E(-\vex{k})-\mu)}{\left[E(\vex{k})+E(-\vex{k})-2\mu\right]^3}-\frac{1}{2}\frac{n_F'(E(\vex{k})-\mu)+n_F'(-E(-\vex{k})-\mu)}{\left[E(\vex{k})+E(-\vex{k})-2\mu\right]^2}.
	\end{align}
    \end{widetext}
	
	To understand how SC is developed, we examine the $\vex{k}$-dependent contribution to $a_1(T)$ and $a_2(T)$ for different $l$. We define $\chi_1(\vex{k})$ and $\chi_2(\vex{k})$ via $a_1=\sum_{\vex{k}}'\chi_1(\vex{k})$ and $a_2=\sum_{\vex{k}}'\chi_2(\vex{k})$. In Fig.~\ref{Fig:chi_k}, we plot $\chi_1(\vex{k})$ and $\chi_2(\vex{k})$ for $l=1,3,5$ with the Fermi level right at VHS. The results show that the peaks of $\chi_a(\vex{k})$ do not simply track the Fermi surface. The results here reveal the complicated nature of intravalley SC owing to $E(\vex{k})\neq E(-\vex{k})$. Another interesting technical finding is that the $p$-wave and the $h$-wave cases yield the same $a_1$ and $a_2$, while the $\chi_1$'s and $\chi_2$'s have very different $\vex{k}$-dependence.

\section{Angular momentum mixing terms}\label{App:mixing}

In a system with three-fold rotational symmetry, there are additional symmetry-allowed terms in the pairing interactions, which can be expressed by
\begin{align}
    \nonumber\hat{H}_{\text{mixing}}\approx&-\frac{1}{2\mathcal{A}}\sum_{\vex{k},\vex{k}'}\sum_{l}\sum_{n\neq 0}\tilde{W}_{l,l-3n}e^{il(\varphi_{\vex{k}}-\varphi_{\vex{k}})}e^{i3n\varphi_{\vex{k}'}}\\
    &\hspace{2.8cm}\times\psi^{\dagger}_{\vex{k}}\psi^{\dagger}_{-\vex{k}}\psi_{-\vex{k}'}\psi_{\vex{k}'}\\
    =&-\frac{1}{2\mathcal{A}}\sum_{\vex{k},\vex{k}'}\sum_{l}\sum_{n\neq 0}\tilde{W}_{l,l-3n}\mathcal{C}^{\dagger}_{l}(\vex{k})\mathcal{C}_{l-3n}(\vex{k}'),
\end{align}
where $\tilde{W}_{l,l-3n}$ is the coefficient of for mixing between $l$ and $l-3n$, $\varphi_{\vex{k}}$ is the angle of $\vex{k}$ relative to the $x$-axis, and $\mathcal{C}_L(\vex{k})=e^{-iL\varphi_{\vex{k}}}\psi_{-\vex{k}}\psi_{\vex{k}}$ annihilates a Cooper pair (with electron momenta $\vex{k}$ and $-\vex{k}$) that carries angular momentum $L$. The interaction $\hat{H}_{\text{mixing}}$ mixes Cooper pairs with angular momenta differed by $\Delta l=3n$. The leading effect is $\Delta l=\pm 3$ ($n=1$), inducing singlet-triplet mixing. In our case of spin-polarized, valley-polarized SC, singlet pairing is not allowed. As a result, the $\Delta l=\pm 3$ process is irrelevant. The same is true for other odd number $n$'s. For an even number $n\neq 0$, mixing between two triplet pairings is possible. However, we expect that such higher-order processes (e.g., $\Delta l=\pm 6$) do not have any physical significance.

For the static screened Coulomb interaction within RPA, the dominant pairing is in the $p$-wave triplet pairing, and the subleading $g$-wave singlet pairing is at least an order of magnitude smaller [Table~\ref{tab:W_l}]. Other interactions, $s$-, $d$-, $f$-, and $h$-wave scatterings, are all repulsive and impossible to facilitate SC. As we discussed in the previous paragraph, the singlet-triplet pairing is irrelevant to the triplet-only situation here. The leading triplet-triplet mixing, the hybridization between the $p$- and $h$-wave, is also irrelevant here because the $h$-wave scattering is repulsive. Thus, we conclude that the angular momentum mixing terms allowed by the three-fold rotational symmetry are irrelevant to our study.


\begin{thebibliography}{80}
\bibitem [{pai()}]{pairing_int}%
	\BibitemOpen
	\href@noop {} {}\bibinfo {note} {The pairing interaction does not need to be
		completely $|\vex{k}|$, $|\vex{k}'|$ independent as the flat band region is
		roughly $|\vex{k}|a<0.1$ as shown in
		Fig.~\ref{Fig:Main_results}(b)}\BibitemShut {NoStop}%
\end{thebibliography}

%

\end{document}